\documentclass[10pt,prl,aps,twocolumn,superscriptaddress,preprintnumbers]{revtex4-1}

\usepackage[utf8]{inputenc}
\usepackage{epigraph}
\usepackage{amsmath}
\usepackage{wrapfig}
\usepackage{epsfig}
\usepackage{amssymb}
\usepackage{graphicx}
\usepackage{slashed}
\usepackage{xcolor}
\usepackage{comment}
\usepackage{natbib}
\usepackage{enumitem}
\usepackage{tikz-cd,tikz}
\usepackage{float}
\usepackage{array,adjustbox,booktabs}
\usepackage{subcaption}
\usepackage{amsfonts}
\usepackage{physics}
\usepackage{listings}
\usepackage{calc}
\usepackage{soul}
\usepackage[normalem]{ulem}

\newcommand\beqa{\begin{eqnarray}}
\newcommand\eeqa{\end{eqnarray}}
\newcommand{\beq}{\begin{eqnarray}}
\newcommand{\eeq}{\end{eqnarray}}

\newcommand{\eq}[1]{(\ref{#1})}

\newcommand\mathstyle{\lstset{
language=Mathematica,
basicstyle={\scriptsize\def\fvm@Scale{.5}\fontfamily{fvm}\selectfont},
otherkeywords={self},
keywordstyle=\ttb\scriptsize\color{deepblue},
emph={MyClass,__init__},
emphstyle=\ttb\color{deepred},
backgroundcolor=\color{pink!20!white},
stringstyle=\color{deepgreen},
commentstyle=\color{SkyBlue3!70!PaleGreen4},
frame=tb,
showstringspaces=false
}}

\newwrite\todofile
\immediate\openout\todofile=\jobname.tdo

\newcounter{todocounter}

\newcommand{\printtodos}{
        \section*{To-Do List}
        \immediate\closeout\todofile
        \input{\jobname.tdo}
}
\lstnewenvironment{mathematica}[1][]
{
\mathstyle
\lstset{#1}
}
{}

\newcolumntype{C}[1]{>{\centering\arraybackslash}m{#1}}

\newcommand{\ii}{i}

\newcommand{\algpsu}{\mathfrak{psu}}

\newcommand{\bQ}{\mathbf{Q}}
\newcommand{\bP}{\mathbf{P}}

\usepackage[hidelinks,colorlinks=true]{hyperref}

\begin{document}
\title{Short Strings in Three-Dimensional Anti-de Sitter Space: \\ from Weak to Strong Coupling}

\author{Simon Ekhammar}
\email{simon.ekhammar@kcl.ac.uk}
\affiliation{Department of Mathematics, King’s College London,
Strand WC2R\,2LS, London, UK}
\affiliation{Department of Physics and Astronomy,
     Uppsala University,
     Box 516,
     SE-751\,20 Uppsala,
     Sweden}
\author{Nikolay Gromov}
\email{nikolay.gromov@kcl.ac.uk}
\affiliation{Department of Mathematics, King’s College London,
Strand WC2R\,2LS, London, UK}
\author{Bogdan Stefa\'nski, jr.}
 \email{bogdan.stefanski.1@citystgeorges.ac.uk}
\affiliation{Centre for Mathematics Innovation, City St George's, University of London, Northampton Square, EC1V\,0HB, London, UK}
\author{Charles Thull}
 \email{charles.thull@citystgeorges.ac.uk}
\affiliation{Centre for Mathematics Innovation, City St George's, University of London, Northampton Square, EC1V\,0HB, London, UK}

\begin{abstract}
We numerically solve the conjectured Quantum Spectral Curve for strings on AdS$_3\times$S$^3\times$T$^4$ with R--R charge from weak to strong coupling. At strong coupling, the spectrum organises into flat-space string mass levels with universal square-root scaling in the string tension at leading order, and additional Kaluza-Klein fine-splitting at subleading order. At weak coupling, the energies are determined by a nearest-neighbour Bethe Ansatz, with universal subleading corrections that are suppressed at large volume.
\end{abstract}

\maketitle
\section{Introduction}
The AdS/CFT correspondence~\cite{Maldacena:1997re} is an exact duality relating free strings on anti-de Sitter (AdS) space to planar conformal field theory (CFT). In principle, solving string theory in AdS provides non-perturbative access to strongly-coupled CFTs. However, first-principle quantisation of Ramond-Neveu-Schwarz strings in many anti-de Sitter backgrounds remains an outstanding problem, since Ramond-Ramond (R--R) vertex operators introduce non-local worldsheet interactions. Remarkably, in certain highly symmetric backgrounds, Green–Schwarz string theory is integrable \cite{Bena:2003wd}, suggesting that the full spectrum can be encoded into a set of integrability-based equations. This hope has been realised in AdS$_5\times$S$^5$ and AdS$_4\times \mathbb{CP}^{3}$ through the Quantum Spectral Curve (QSC) formalism~\cite{Gromov:2013pga,Cavaglia:2014exa}.

\begin{figure}
    \centering
    \includegraphics[width=\linewidth]{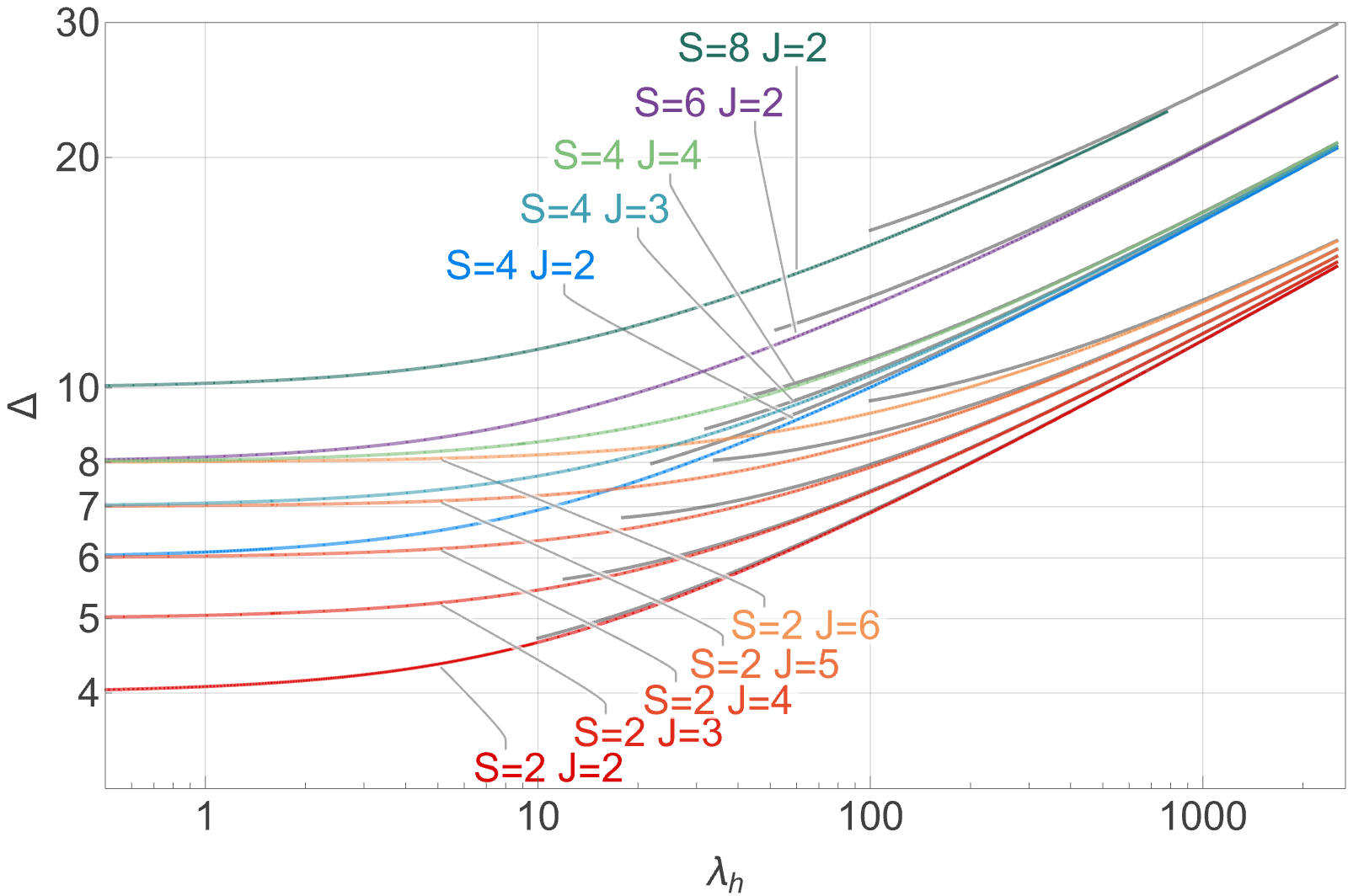}
    \caption{The numerical QSC spectrum for a wide range of couplings, $\lambda_h\equiv 16\pi^2 h^2$, and for different states labelled by their spin, $S$, and R-charge, $J$. A stringy spectrum emerges at strong coupling, with energy levels in agreement with flat-space string theory together with additional KK-modes for the two lowest mass-levels. The gray lines show the strong coupling prediction \eq{eq:DeltaStrong}.}
    \label{fig:NumericalResult}
\end{figure}

Inspired by this higher-dimensional success, a QSC for free strings on AdS$_3\times$S$^3\times$T$^4$ supported by pure R--R charges was conjectured in~\cite{Cavaglia:2021eqr,Ekhammar:2021pys}. Compared to its higher-dimensional cousins, this model presents many additional conceptual and technical challenges~\cite{Cavaglia:2022xld,Ekhammar:2024kzp}. In this Letter, we overcome these difficulties to find non-trivial solutions of the QSC numerically and, for the first time, predict the AdS$_3$ string spectrum from weak to strong coupling—a goal that has driven the integrability programme since its inception more than 15 years ago~\cite{Babichenko:2009dk}. Our results, summarised in Figure~\ref{fig:NumericalResult}, show the conformal dimensions of low-lying operators throughout the full coupling range in terms of $\lambda_h\equiv 16\pi^2 h^2$, with $h$ the integrability coupling constant.\footnote{The zero-coupling values $\Delta_0$ shown as the x-intercept, reflect our particular choice of operators and should not be taken as exhaustive: other values, including half-integer ones, will exist for other operator choices.}

\paragraph{AdS$_3$ and Holography.}
Strings on AdS$_3\times$S$^3\times$T$^4$ occupy a distinguished place in holography. Through the AdS/CFT correspondence, they are expected to describe the near-horizon D1-D5 system~\cite{Maldacena:1997re}, which has played a central role in microscopic black-hole entropy counting. Moreover, two-dimensional CFTs are much more constrained than their higher-dimensional counterparts, raising the prospect of a particularly sharp realisation of holography. At the same time, AdS$_3$ backgrounds exhibit features absent in higher dimensions, including long strings, continuous spectra, and a multi-dimensional moduli space acted on by a large U-duality group, making them a fertile testing ground for holographic ideas, see for example~\cite{Giveon:1998ns,Seiberg:1999xz,Larsen:1999uk,Maldacena:2000hw,Aharony:2024fid}.

For AdS$_3$ backgrounds supported purely by Neveu-Schwarz-Neveu-Schwarz (NS--NS) flux, the string spectrum can be determined using Wess-Zumino-Witten (WZW) methods~\cite{Maldacena:2000hw}. Additionally, at minimal NS--NS flux the hybrid formalism~\cite{Berkovits:1999im} yields an exact match with the symmetric-orbifold CFT~\cite{Gaberdiel:2018rqv,Eberhardt:2018ouy,Dei:2019osr}. However, away from these special loci, the duality is expected to exhibit significantly richer dynamics, especially along continuous ’t~Hooft-like coupling directions that are believed to realise weak-strong holography, as in higher-dimensional AdS/CFT examples.  
Progress in this regime therefore hinges on the availability of genuinely quantitative methods, such as the QSC framework. 

\paragraph{Weak Coupling and the CFT Regime.}

At weak coupling $h \ll 1$, the anomalous dimensions for our QSC solutions $\Delta$ satisfy 
\begin{equation}\label{eq:weak-coupling}
\Delta = \Delta_0 + h^2\Delta_1 + \mathcal{O}(h^3)\,,
\end{equation}
with $\Delta_0 \in \mathbb{Z}_{>0}$. The leading corrections $\Delta_1$ coincide with the spectrum of an integrable nearest-neighbour $\mathfrak{sl}_2$ spin-chain, in agreement with the structure for the weakly-coupled planar description of the CFT proposed in~\cite{OhlssonSax:2014jtq}. Our results extend previous analytical calculations~\cite{Cavaglia:2022xld} to operators with larger charges and uncover novel wrapping corrections at $\mathcal{O}(h^3)$, providing the first controlled access to finite-size effects at weak coupling. These findings offer concrete spectral CFT data in a region of moduli space far removed from the symmetric orbifold point.

\paragraph{A Stringy Spectrum.}

At strong coupling $\lambda_h \gg 1$, the integrability coupling is expected to be related to the inverse string tension, $\alpha'$, as $\sqrt{\lambda_h} \simeq R_{\text{AdS}}^2/\alpha'$, with $R_{\text{AdS}}$ the radius of AdS$_3$, \textit{cf}~\cite{OhlssonSax:2018hgc}. This relation suggests that the strong-coupling regime may be viewed as a flat-space limit, in which the spectrum should organize itself into Regge trajectories labelled by mass levels $\Delta^2 \sim 4\delta/\alpha' \simeq 4\sqrt{\lambda_h}\,\delta$ for positive integer $\delta$ (in $R_{\text{AdS}}=1$ units), where $\delta$ counts the total string oscillator excitations' level. The compact S$^3$ introduces Kaluza-Klein (KK) modes, splitting each Regge trajectory at subleading order in $\lambda_h$ into a tower of states with different spherical angular momenta.

The QSC is a symmetry-based construction following from $\mathfrak{psu}(1,1|2)^{\oplus 2}$ representation theory and analyticity constraints, and, a priori, it is not obvious that it knows about string theory. Still, our solutions reveal the emergence of Regge trajectories, KK towers, and the characteristic $\lambda_h^{1/4}$ scaling expected for strings on AdS$_3\times$S$^3\times$T$^4$. Our numerical results, summarised in Figure~\ref{fig:NumericalResult}, reveal precisely this structure at strong coupling, with states clustering into string levels characterised by
\begin{equation}\label{eq:FlatSpace}
\Delta = 2\sqrt{\delta}\,\lambda_h^{\frac{1}{4}} + E_0 + E_1\,\lambda_h^{-\frac{1}{4}} + E_2\,\lambda_h^{-\frac{2}{4}} + \mathcal{O}(\lambda_h^{-\frac{3}{4}})\,,
\end{equation}
where all $E_n$ coefficients, except $E_0$ which we found to be $0$, depend on the angular momentum in S$^3$. This provides strong support for the conjecture~\cite{Cavaglia:2021eqr,Ekhammar:2021pys} that the QSC captures the quantum spectrum of free strings in this background.

In what follows, we present the numerical methods underlying these results and make quantitative comparisons with semi-classical strings at strong coupling and the Asymptotic Bethe Ansatz (ABA) at weak coupling. The QSC provides for the first time access to the full coupling range, furnishing concrete spectral data for precision holography in AdS$_3$.

\section{The AdS$_3$ Quantum Spectral Curve}

The starting point for the QSC is the superalgebra $\mathfrak{psu}(1,1|2)^{\oplus 2}$, reflecting the left--right symmetry of a non-chiral two-dimensional CFT and generating the super-isometries of AdS$_3 \times$S$^3$. We distinguish the right-moving copy by using
dotted indices. This algebra has two spacetime charges, the conformal dimension $\Delta$ and the spin $S$, as well as two R-symmetry charges $J$ and $K$. The role of the QSC is to lift this classical symmetry algebra to a quantum deformation.

This is achieved through a supersymmetric Q-system~\cite{Kazakov:2007fy,Tsuboi:2009ud,Kazakov:2015efa}, which promotes relations amongst characters to Hirota-like finite difference equations. The Q-system comprises functions of a spectral parameter $u$ whose large-$u$ asymptotics encode the global charges, whilst their analytic structure—controlled by the coupling $h$—encodes the infinite tower of integrable charges. Each $\mathfrak{psu}(1,1|2)$ Q-system contains four distinguished Q-functions, $\mathbf{P}_{a},\mathbf{Q}_{k}$ with $a,k=1,2$, from which all other Q-functions can be constructed. We recall the complete structure in Section~\ref{SM:AdS3QSC} of the Supplemental Material.

The quantum deformation is specified by demanding polynomial asymptotics
\begin{align}
    \bP_{a} \sim u^{M_{a}}\,,\,\ \bQ_{k} \sim u^{\hat{M}_{k}}\,,\,\ \bP_{\dot{a}} \sim u^{M_{\dot{a}}}\,,\,\ \bQ_{\dot{k}} \sim u^{\hat{M}_{\dot{k}}}\,,
\end{align}
where $M,\hat{M}$ are linear combinations of the classical charges $\Delta,S,J,K$~\cite{Cavaglia:2021eqr}. Higher integrable charges are encoded in $1/u$ corrections to these asymptotics. However, the lift from classical to quantum algebra is not unique — different analytic properties yield different integrable models. For AdS$_3\times$S$^3\times$T$^4$, the analytic structure was conjectured in~\cite{Ekhammar:2021pys,Cavaglia:2021eqr}: the $\mathbf{P}$-functions have a short branch cut on $[-2h,2h]$, introducing the coupling~$h$, whilst the $\mathbf{Q}$-functions possess an infinite tower of cuts spaced by $i$ extending into the lower half-plane from the real axis. The two Q-systems are then related by gluing along $[-2h,2h]$:
\begin{equation}\label{eq:gluing}
    \mathbf{Q}_{k}(u+i 0) =  G_{k}{}^{\dot {p}}\,\overline{\mathbf{Q}}_{\dot{p}}(u-i 0)\,,
    \quad
    G = \begin{pmatrix}
        0 & i\alpha\\
        -i/\alpha & 0
    \end{pmatrix}\,,
\end{equation}
where $\alpha$ is a real function of the coupling and $\overline{\mathbf{Q}}_{\dot{l}}$ denotes complex conjugation. These analytic properties, together with the algebraic Q-system relations, fully quantise the model and determine its spectrum.

In this Letter we focus on solutions with quantum numbers
$\left.\Delta\right|_{h=0}=J+S$ and $K=0$
and parity symmetry $u\leftrightarrow- u$. The subset of these states  with $J=2$ was previously studied at weak coupling in~\cite{Cavaglia:2022xld}.

\subsection{QSC at weak coupling}

We begin in the weak-coupling regime, where the QSC can be solved analytically and its analytic structure can be elucidated. This will inform the subsequent numerical analysis at strong coupling.

The key technical feature distinguishing AdS$_3$ from higher-dimensional cases is that the branch points at $\pm 2h$ are logarithmic rather than quadratic~\cite{Borsato:2013hoa}, reflecting the presence of gapless worldsheet modes. This necessitates a double-sum Ansatz~\cite{Cavaglia:2022xld}:
\begin{equation}\label{eq:WeakExpansion}
    \bP_{a} = \sum_{k=0}^{\infty}\sum_{m=0}^{k} h^k l^{m}\bP^{(k,m)}_{a}\,,
\end{equation}
where $l=\frac{i}{2\pi}\log\left(\frac{x-1}{x+1}\right)$ and $\bP_{a}^{(k,m)}$ are Laurent polynomials in the Zhukovsky variable \begin{equation}\label{eq:Zhukovsky}
    x(u)=\frac{1}{2h}\left(u+\sqrt{u^2-4h^2}\right).
\end{equation} Solving the Q-system relations subject to the gluing condition \eqref{eq:gluing} and extracting dimensions from large-$u$ asymptotics, we extend previous $J=2$ results~\cite{Cavaglia:2022xld} to the full $J=3$ sector (computed for $S=2,\ldots,12$, but expected to hold for general $S$):
\begin{align}\label{eq:L3gamma}
    \Delta(J{=}3,S) &=J+S+ 8 \mathbf{S}_1\,h^2+\frac{512}{63\pi}\mathbf{S}_1^2\,h^3\nonumber\\ 
    &\hspace{6pt}+ \left(\gamma_{\mathcal{N}=4}^{(4),J=3}-\frac{8192}{525\pi^2}\mathbf{S}_{1}^{3}\right) h^{4}+\mathcal{O}(h^5),
\end{align}
with harmonic sums $\mathbf{S}_k=\sum_{n=1}^{S/2}n^{-k}$ and $\gamma_{\mathcal{N}=4}^{(4),J=3}=-8\mathbf{S}_3-4\mathbf{S}_1\mathbf{S}_2$~\cite{Kotikov:2007cy}. For $J=2$, we recover the result of~\cite{Cavaglia:2022xld} and verify its validity at larger values of $S$.

At $\mathcal{O}(h^2)$, these results reproduce the AdS${}_3$ ABA and match the corresponding $\mathcal{N}=4$ SYM dimensions. Intriguingly, for both $J=2,3$ all terms computed match $\mathcal{N}=4$ SYM in the formal limit $1/\pi\to 0$~\cite{Cavaglia:2022xld}. The $\mathcal{O}(h^3)$ term is absent in the ABA, and as argued below, is suppressed as $J\to\infty$; we would thus interpret it as a wrapping correction.  Remarkably, the coefficient $\gamma_3$ of $h^3$ is proportional to $(\gamma_2)^2$ (where $\gamma_2$ is the $h^2$ coefficient) with an $S$-independent rational prefactor. Combining these analytic results for arbitrary $S$ with numerical fits (as described below) at $S=2,\,J=4,5,\ldots,8$ and $S=4,\,J=4$, we find a universal expression
\begin{equation}\label{eq:WeakCoupling}
    \gamma_3= \frac{2(J+1)}{(2J+1)(2J+3)\pi} (\gamma_{2})^2\,,
\end{equation}
which we conjecture to hold for general $J$ and $S$. The existence of such an unexpected relation suggests that there may be a modified ABA capable of capturing small-$h$ corrections rather than large-volume effects, analogous to the $\mathcal{N}=4$ case~\cite{Beisert:2005fw}. Such a formulation would provide analytical weak-coupling starting points essential for systematic numerical studies and warrants future investigation.

\section{Numerical Approach}
To reach finite and strong coupling, numerical methods are essential, since no exact analytic expression $\Delta(h)$ in terms of elementary functions is likely to exist, as evidenced by the non-trivial weak coupling expansion. In AdS$_5$ and AdS$_4$, where branch cuts are quadratic, one parametrizes the $\mathbf{P}$-functions as
\begin{equation}\label{eq:OldNumericsAnsatz}
    \bP_{a} \simeq x^{M_{a}}\sum_{k=0}^{\text{cut-off}} \frac{c_{a,k}}{x^{2k}}\,,
\end{equation}
where the $x^{-2k}$ basis exploits parity symmetry. The Zhukovsky variable $x$ resolves the quadratic branch cut singularity, yielding exponential convergence even at $|x|=1$ (i.e., on the cut in $u$). However, in AdS$_3$ the logarithmic branch-cut structure means $x$ no longer removes the branch-points. For example, the logarithm $l$ from \eqref{eq:WeakExpansion} expands as $l\propto \sum_{k\,\mathrm{odd}} (kx^{k})^{-1}$, exhibiting only power-law convergence $\sim 1/k$. This severely limits the numerical efficiency of this basis.

Standard convergence acceleration addresses this by using basis functions that resum such slowly-converging series. Motivated by the logarithmic structure in \eqref{eq:WeakExpansion}, we introduce the polylogarithm combinations
\begin{align}
    L_n(x)=\sum_{k=1}^{n}a_{k,n}\mathrm{Li}_k\left(x^{-2}\right)=x^{-2n}+\mathcal{O}(x^{-2n-2}),
 \end{align} 
where the right-hand side implicitly defines the coefficients $a_{k,n}$. With the modified ansatz
\begin{align}\label{eq:NumericalNew}
     \bP_a = x^{M_a}\sum_{k=0}^{\text{cut-off}} c_{a,k} L_k(x)\,,
 \end{align} 
we achieve dramatically improved performance, reaching reliable results up to $h\sim 4$ ($\lambda_h \sim 2500$), a substantial improvement over the $h<0.1$ limit of previous methods~\cite{Cavaglia:2022xld}, which we reproduce in the overlap region.

\paragraph{The numerical algorithm.} With the ansatz \eqref{eq:NumericalNew}, we solve the QSC following~\cite{Gromov:2015wca}. We first solve the Q-system relations and then evaluate the gluing condition~\eqref{eq:gluing} on the cut $[-2h,2h]$. For a set of sampling points $I\subset[-2h,2h]$, we minimise the cost function
\begin{equation}\label{eq:cost}
F(\{c_{a,n},c_{\dot{a},n}\},\Delta)=\sum_{u\in I}\left|\mathbf{Q}_k(u+i 0)-G_k{}^{\dot{l}}\,\overline{\mathbf{Q}}_{\dot{l}}(u-i 0)\right|^2\,,
\end{equation}
which vanishes for exact solutions. We use a Newton-type method to optimise the coefficients $c_{a,n}$, $c_{\dot{a},n}$, the dimension $\Delta$ and the gluing matrix parameter $\alpha$.

A key refinement is the choice of sampling points $I$. Rather than the standard Chebyshev nodes, we concentrate points near the ends of branch-cuts $\pm 2h$ via
$
    u_i = 2h\,\mathrm{sign}(t_i)\left[1-(1-|t_i|)^\beta\right],
$
where $\{t_i\}$ are roots of the Chebyshev polynomial $T_n(t)$ and  $\beta=5/3$. This choice provides good numerical stability and improves the resolution of the singular behaviour where the gluing condition is most sensitive.

\section{The Spectrum At Strong Coupling}
We study two series of states (Figure~\ref{fig:NumericalResult}): $(J{=}2,\, S{=}2,4,6,8)$ and the KK tower $(S{=}2,\, J{=}2,\ldots,8)$, as well as states with $S=4,\;J{=}3,4$. Our numerical results are presented in Supplemental Material~\ref{SM:data}.

Fitting our numerical data to a series in $\lambda_h^{-1/4}$, we find excellent agreement with~\eqref{eq:FlatSpace} for $\delta= S/2$ and $E_0 = 0$. This is highly non-trivial: different QSCs exhibit distinct strong-coupling asymptotics, and the QSC construction itself contains, a priori, no string theory input. Moreover, we find $E_{2}\neq 0$, a novel feature absent in higher-dimensional examples. 

To extract subleading coefficients, we use the fitting procedure of~\cite{Gromov:2023hzc}, where following $\mathcal{N}=4$ SYM, we assume terms $\lambda_h^{-a/2}$ with $a>1$ are absent. Our results strongly suggest that, up to an overall $1/\sqrt{S}$ factor, the $\lambda_h^{-1/4}$ coefficient depends only on $J^2,S,S^2$. Assuming this structure, we find, with an uncertainty in last digit,
\begin{equation}\label{eq:lambda14}
    4\sqrt{2 S}\,\Delta\big|_{\lambda_h^{-1/4}} = 2.00000\, J^2 - 3.9995\, S + 2.9998\, S^2\,.
\end{equation}
The $\lambda_h^{-1/2}$ term is simpler, scaling linearly with $J$:
\begin{equation}\label{eq:lambda12}
    \Delta\big|_{\lambda_h^{-1/2}} = 0.50002\, J\,.
\end{equation}
These numerical values strongly suggest integer coefficients, as in higher dimensions. We therefore conjecture the exact expansion
\begin{equation}\label{eq:DeltaStrong}
    \Delta = \sqrt{2\,S}\,\lambda_h^{1/4}+\frac{2 J^2-4\,S+3\,S^2}{4 \lambda_h^{1/4}\sqrt{2 \,S}}+\frac{J}{2\lambda_h^{1/2}} + \dots\,.
\end{equation}
This resembles the AdS$_5\times$S$^5$ short string spectrum, with the key novelty being the $\lambda_h^{-1/2}$ term. Figure~\ref{fig:NumericalResult} shows excellent visual agreement between \eqref{eq:DeltaStrong} and our numerical data.

The AdS Virasoro-Shapiro amplitude \cite{Alday:2023mvu} was recently extended to AdS$_3$~\cite{Chester:2024wnb,Jiang:2025oar,Jiang:2026xnh}, leading to results up to $\mathcal{O}(\lambda^{-1/4})$ for related but distinct states (see~\ref{sec:comp-vs}). They find an expression similar to \eqref{eq:DeltaStrong}, differing only by $-4S\to-2S$ in the $\lambda^{-1/4}$ coefficient. Since the states differ, no direct comparison is possible; moreover, the coefficient can be shifted by a state-independent reparametrization $\lambda_h(\lambda)$. The next-order term at $\lambda_h^{-3/4}$, which we provide below, will be essential for a meaningful comparison with independent approaches.

Extracting the $\lambda_h^{-3/4}$ coefficient is considerably more challenging, as numerical precision degrades rapidly for subleading terms. Motivated by $\mathcal{N}=4$ SYM, where states on the same conformal Regge trajectory as a BPS operator exhibit a particular structure, we assume that the coefficient takes the form $(2S)^{-3/2}$ times a linear combination of $J^4,J^2S,J^2S^2,S^2,S^3,S^4$ \cite{Gromov:2011bz}. We emphasise that this is an assumption which requires future justification—it is known to fail for generic states in $\mathcal{N}=4$ SYM, which contain non-polynomial dependence on charges due to mixing~\cite{Ekhammar:2024rfj}. Fitting all our states (excluding $S=8$, for which we have not reached $h=4$), we find
\begin{equation}
\begin{split}
32 (2 S)^{3/2}\Delta\big|_{\lambda_h^{-3/4}} &= -3.9991\, J^4+19.9\, S^2 J^2-15.8\, S J^2\\
&\quad-21.0006\, S^4-1.98\, S^3-23.5\, S^2\;.
\end{split}
\end{equation}
Some coefficients appear reliably integer, whilst others do not. We therefore propose
\begin{equation}\label{eq:Lambda34}
\begin{split}
    &\Delta|_{\lambda_h^{-3/4}} = \\
    &\frac{20\,J^2\,S^2-4J^4-21 S^4+c_1\,J^2 S+c_{2} S^2+c_3 S^3}{32 \left(2S\right)^{3/2}}\,,
\end{split}
\end{equation}
leaving $c_1$, $c_2$, $c_3$ undetermined. Our data suggest that $c_1$ and $c_3$ may also be integers. If one further assumes that any deviation from integer values arises from transcendental terms proportional to $\zeta_3$, as occurs in $\mathcal{N}=4$, the best fit yields (with some uncertainty)
\begin{equation}
    c_1 \simeq -16\,,
    \quad
    c_{2} \simeq -2\,,
    \quad
    c_{3} \simeq 92-96\zeta_3\,.
\end{equation}
Extending the Virasoro-Shapiro calculations of~\cite{Chester:2024wnb,Jiang:2025oar,Jiang:2026xnh} to this order would provide valuable insight into which coefficients are integers and which contain transcendental numbers.

\paragraph{Classical String Comparison.}
Beyond comparing to flat-space string expectations, we can test our results against classical strings in curved AdS space. We compare against the folded string solution~\cite{Gubser:2002tv,Frolov:2002av}, which is the lowest-energy classical configuration with the same quantum numbers and parity symmetry as our states. Whilst a classical solution describes a macroscopic string with charges scaling as $S,J\sim \sqrt{\lambda}_h$, we can re-expand the classical expression~\cite{Gromov:2011bz} to compare with our short-string results:
\begin{equation}\label{eq:ClassicalExpanded}
    \Delta_{\text{cl}} = \sqrt{2 S}\,\lambda_h^{\frac{1}{4}} +\frac{2 J^2+3 S^2}{4 \sqrt{2 S}\, \lambda_h^{1/4}}-\frac{4 J^4-20 J^2 S^2+21 S^4}{32 \left(2 S\right)^{3/2} \lambda_h ^{3/4} }+\dots\,.
\end{equation}
This yields perfect agreement with the
terms with maximal powers of $S$ and $J$ at  $\lambda_h^{1/4},\;\lambda_h^{-1/4}$
in \eqref{eq:DeltaStrong} and
$\lambda_h^{-3/4}$ in \eqref{eq:Lambda34}.
This provides yet another non-trivial confirmation that the QSC captures quantum string dynamics in AdS.
The lower order terms from \eqref{eq:DeltaStrong} and \eqref{eq:Lambda34} absent in \eqref{eq:ClassicalExpanded} represent quantum corrections to the classical energy.
In particular, the $\lambda_h^{-1/2}$ term should be captured by a one-loop calculation. It would be interesting to revisit previous results~\cite{Beccaria:2012kb,Beccaria:2012pm,Forini:2012bb} in light of our findings.

\section{Conclusions and Outlook}
We have found non-trivial solutions of the conjectured R--R AdS$_3\times$S$^3\times$T$^4$ QSC from weak to strong coupling, determining for the first time the spectrum of classes of short strings across a large coupling range. Our findings reveal a rich structure bridging integrable spin-chain physics at weak coupling and stringy Regge trajectories at strong coupling.

At weak coupling, the spectrum starts from a nearest-neighbour $\mathfrak{sl}_2$ integrable spin-chain~\cite{OhlssonSax:2014jtq}, mirroring $\mathcal{N}=4$ SYM and ABJM. Novel features emerge at higher orders: odd powers of the coupling reminiscent of BFKL physics~\cite{Ekhammar:2024neh}, and inverse powers of $\pi$ suggestive of infrared effects from massless modes. Understanding their precise origin — for instance, as infrared bubble effects within the perturbative framework of~\cite{OhlssonSax:2014jtq} — may illuminate a new formulation of the CFT near the weakly-coupled R--R point in moduli space.

At strong coupling, short strings organise into flat-space Regge trajectories splitting into S$^3$ KK-towers, with non-vanishing $\lambda_h^{-1/2}$ corrections consistent with infrared dynamics of massless modes~\cite{Roiban:2009aa}.\footnote{We note that the presence of terms at this order in the strong-coupling expansion is not entirely without precedent in related contexts; the reader may find it instructive to consult the discussion surrounding eq. (4.10) of~\cite{Arutyunov:2005hd}, where a careful analysis of the $\mathfrak{su}(1|1)$ sector of AdS$_5\times$S$^5$ reveals a broadly analogous structure, the precise interpretation of which in that setting is discussed in subsequent sections by the authors.} Remarkably, this stringy structure emerges from a purely symmetry-based construction, providing strong evidence for the QSC conjecture~\cite{Cavaglia:2021eqr,Ekhammar:2021pys}.

Our methods open several compelling directions. States with massless excitations are accessible with modified analytic properties and $h^1$ scaling, which it would be interesting to compare with related results obtained using the Thermodynamic Bethe Ansatz \cite{Frolov:2021bwp,Brollo:2025rgp}. Adapting our approach to states studied via Virasoro-Shapiro amplitudes~\cite{Chester:2024wnb,Jiang:2025oar,Jiang:2026xnh} (see Supplemental Material~\ref{sec:comp-vs}) would provide a direct test of the QSC. Similarly, adapting our methods to the computation of the Hagedorn temperature will allow comparison against effective string theory results, successful in higher dimensions to several subleading orders~\cite{Harmark:2021qma,Ekhammar:2023glu,Ekhammar:2023cuj}. Mixed-flux backgrounds~\cite{Hoare:2013pma,Lloyd:2014bsa,Hoare:2013lja}, where the $h\to 0$ limit should reproduce~\cite{OhlssonSax:2018hgc} the WZW spectrum~\cite{Maldacena:2000hw}, offer another important setting, cf. the recent proposal for Thermodynamic Bethe Ansatz equations \cite{Frolov:2025tda}. Finally, extending our numerical approach to AdS$_3\times$S$^3\times$S$^3\times$S$^1$~\cite{Babichenko:2009dk,OhlssonSax:2011ms,Cavaglia:2025icd,Chernikov:2025jko} is a further interesting direction. 

\begin{acknowledgments}
\section{acknowledgments}

We are grateful to A.~Cavaglià, F.~Chernikov, P.~Ryan, A~Torrielli, and D.~Volin for collaboration on related topics as well as to 
 B.~Basso, S.~Chester, S.~Frolov, A.~Georgoudis, D.~Polvara, A.~Sfondrini, A.~Tseytlin and D.-l.~Zhong for fruitful discussions. 

The work of S.E. was conducted with funding awarded
by the Swedish Research Council, Vetenskapsrådet, grant VR 2024-00598. The work of N.G was
supported by the European Research Council (ERC) under the European Union’s Horizon 2020 research and innovation program – 60 – (grant agreement No. 865075) EXACTC. 
B.S. and C.T. acknowledge funding support
from an STFC Consolidated Grant ‘Theoretical Particle Physics at City, University of London' ST/T000716/1.
The work of N.G. was supported by the Science Technology \& Facilities Council under the grants ST/P000258/1 and ST/X000753/1.
\end{acknowledgments}

\bibliography{ref}
\onecolumngrid
\newpage
\setcounter{equation}{0} 
\renewcommand{\theequation}{S\arabic{equation}} 
\setcounter{secnumdepth}{2}

\section{A Brief Reminder of the AdS$_3$ QSC}\label{SM:AdS3QSC}
In this appendix, we recap the general formulas for the AdS$_3$ QSC and give details on the algorithm for the perturbative solution for the particular states considered in this letter.

The AdS$_3$ QSC consists of two $\algpsu(1,1|2)$ Q-systems connected through analytic continuation. Each Q-system consists of 16 Q-functions satisfying QQ-relations \begin{align}
    Q_{a|i}^+-Q_{a|i}^-&=\bP_a\bQ_i, & \bQ_i&=-\bP^a Q_{a|i}^+,\\
    Q_{\dot{a}|\dot{k}}^+-Q_{\dot{a}|\dot{k}}^-&=\bP_{\dot{a}}\bQ_{\dot{k}}, & \bQ_{\dot{k}}&=-\bP^{\dot{a}} Q_{\dot{a}|\dot{k}}^+,
\end{align}
where $f^{\pm}=f(u\pm\frac{i}{2})$. These functions have the asymptotics
\begin{equation}\label{eq:AsymptoticsQSM}
    \bP_{a} \simeq \mathbb{A}_{a} u^{M_{a}}\,,
    \quad
    \bP^{a} \simeq \mathbb{A}^{a} u^{-M_{a}-1}\,,
    \quad
    \bQ_{k} \simeq \mathbb{B}_{k} u^{\hat{M}_{k}}\,,
    \quad
    \bQ^{k} \simeq \mathbb{B}^{k} u^{-\hat{M}_{k}-1},
\end{equation}
\begin{align}
    M_{a} &= \frac{1}{2}\{-J+K-2,J-K\} - \frac{\hat{B}}{2}\,,
    &
    \hat{M}_{k} &= \frac{1}{2}\{\Delta-S,-\Delta+S-2\} + \frac{\hat{B}}{2}\\
    M_{\dot{a}} &= \frac{1}{2}\{-J-K,J+K-2\}-\frac{\check{B}}{2}\,,
    &
    \hat{M}_{\dot{k}} &= \frac{1}{2}\left\{\Delta+S-2,-\Delta-S\right\}+\frac{\check{B}}{2}\,,
\end{align}
and the leading coefficients $\mathbb{A}_a$ are given as 
\begin{align}
    \mathbb{A}_1\mathbb{A}^{1} &= - \mathbb{A}_2 \mathbb{A}^{2} =  -\frac{\ii}{4}\frac{(\Delta-J+K-S)(\Delta+J-K-S+2)}{J-K+1},\\
    \mathbb{A}_{\dot{1}}\mathbb{A}^{\dot1} &= - \mathbb{A}_{\dot2} \mathbb{A}^{\dot 2} =  -\frac{\ii}{4}\frac{(\Delta-J-K+S)(\Delta+J+K+S-2)}{J+K-1},\\
    \mathbb{B}_1\mathbb{B}^{1} &= - \mathbb{B}_2 \mathbb{B}^{2} =  \frac{\ii}{4}\frac{(\Delta-J+K-S)(\Delta+J-K-S+2)}{\Delta-S+1},\\
    \mathbb{B}_{\dot{1}}\mathbb{B}^{\dot1} &= - \mathbb{B}_{\dot2} \mathbb{B}^{\dot 2} =  \frac{\ii}{4}\frac{(\Delta-J-K+S)(\Delta+J+K+S-2)}{\Delta+S-1}.
\end{align} (Following \cite{Cavaglia:2022xld}, we changed $S\rightarrow-S$ and $K\rightarrow-K$ compared to \cite{Ekhammar:2024kzp}.)
The $\bQ$-functions can be found from the $\bP$-functions by solving the Baxter equation \begin{align}
    D_1^- \bQ_k^{++}-D_2\bQ_k+D_1^+\bQ_k^{--}=0,
\end{align} where \begin{align*}
    D_1&=\epsilon_{ab}(\bP^a)^-(\bP^b)^+,\\
    D_2&=\epsilon_{ab}(\bP^a)^{--}(\bP^b)^{++}-\bP_c(\bP^c)^{--}\epsilon_{ab}\bP^a(\bP^b)^{++}.
\end{align*}
As a boundary condition on the Q-system we have the constraint
\begin{equation}\label{eq:detConstraint}
    \det(Q_{a|i})=\det(Q_{\dot a|\dot i})=1.
\end{equation}
On the first Riemann sheet the $\bP$-functions have a short branch cut on the interval $[-2h,2h]$. By consistency, the $\bQ$-functions have on the first Riemann sheet an infinite tower of cuts, with the first one on the real line interval $[-2h,2h]$ and the rest of the tower spaced by $\ii$ and extending into the lower half plane. The $Q_{a|i}$ functions have a similar tower of cuts but shifted by $-\frac{\ii}{2}$ into the lower half-plane.

In this letter, we consider states with $K=0$ and $\Delta(h=0)=J+S$. For these states, we find $\bP_a=\epsilon_{ab}\bP^b$ and the parity relation \begin{align}
    \bP_a(-u)&=\mathbf{g}_a{}^b\bP_b(u),
\end{align} with $\mathbf{g}_{a}{}^{ b}=(-1)^{M_a}\delta_a^b$ and similar for the dotted functions. The analytic continuation of the $\bP$-functions through their first branch cut on a path $\overline{\gamma}$ counter-clockwise around the point $u=-2h$ can then be written in the form \begin{align}
    \bP_a=\mu^R_a{}^{\dot b}(\bP_{\dot{b}})^{\overline{\gamma}}.
\end{align} The matrix $\mu^R$ is constructed as \begin{align}\label{eq:muReg}
    \mu^R_a{}^{\dot b}=-Q_{a|k}(u+\frac{\ii}{2})\epsilon^{kl}G_{l}{}^{\dot m}e^{-\ii\pi\hat{M}_{\dot{m}}}Q_{\dot c|\dot m}\epsilon^{\dot c\dot d}\mathbf{g}_{\dot d}{}^{\dot b},
\end{align} in terms of the gluing matrix \eqref{eq:gluing} and it has no branch cuts on the real line. An analogous equation gives $\mu^R_{\dot a}{}^b$.

\section{Analytic solutions of the QSC}

The QSC can be solved analytically using the coupling $h$ as a small expansion parameter. The algorithm presented here is a refinement of the algorithm developed in \cite{Cavaglia:2022xld}.

For $J=2,3$ the tree-level ($h=0$) $\bP$-functions are monomials in $u$. Using this, it is easy to solve the Baxter equation for $\bQ$ and the QQ-relations for $Q_{a|i}$, followed by applying the determinant constraint to fully fix the tree-level solution.

At the QQ-scale we make the Ansatz \begin{equation}\label{eq:QQscaleAnsatz}
    \bP_a=\mathbb{A}_a u^{M_a}\left(1+\sum_{n=1}^\infty c_{a,n}\frac{h^{2n}}{u^{2n}}\right),
\end{equation} with the coefficients expanding as $c_{a,n}=c_{a,n}^{(0)}+\mathcal{O}(h)$. We use this Ansatz to solve order by order the QQ-relations for the $Q_{a|i}$- and $\bQ$-functions. The asymptotics of the Q-functions and the determinant constraint \eqref{eq:detConstraint} fix most of the integration constants. Using \eqref{eq:muReg} we can then construct $\mu^R(u),\dot{\mu}^R(u)$ with the parameter of the gluing matrix expanding as $\alpha=\alpha^{(0)}h^J+\mathcal{O}(h^{J+1})$. Since these $\mu^R$ matrices do not have a branch cut on the real line, we can change variables to $v=\frac{u}{h}$ and at finite order in $h$ the $\mu^R$ are a finite order polynomial in $v$. Requiring $\mu^R$ to be unambiguous, \textit{viz.} the absence of any regularisation parameter, provides at this stage some additional constraints on the integration constants. We can then compute the matrix $W_a{}^c(v)=\mu^R_a{}^{\dot{b}}\mu_{\dot b}^R{}^c$ which describes the double analytic continuation of the $\bP$-functions \begin{equation}
    \bP_a=W_a{}^b\bP_b^{2\overline{\gamma}}.
\end{equation} By construction $W$ has unit determinant. As the $\bP$-functions should be regular near the branch points $u=\pm 2h$, the trace of $W$ is constrained as \begin{align}
    \textrm{tr}\,W(v=\pm2)=2,
\end{align} such that $W$ approaches the identity matrix near the branch points of $\bP$.

The final constraint in order to fix the solution of the QSC is the analytic continuation. This cannot be imposed using the Ansatz \eqref{eq:QQscaleAnsatz} in the variable $u$. Instead, we make a second Ansatz \eqref{eq:WeakExpansion} using the Zhukovsky variable $x$ \begin{equation}
    \bP_{a} = \sum_{k=0}^{\infty}\sum_{m=0}^{k} h^kl^{m}\bP^{(k,m)}_{a},
\end{equation}
with $l=\frac{\ii}{2\pi}\log\left(\frac{x-1}{x+1}\right)$ and $\bP_{a}^{(k,m)}$ are Laurent polynomials in $x$ of finite degree. Applying the definition of the Zhukovsky variable and expanding the new Ansatz at small $h$ we can relate the coefficients in both Ans{\"a}tze. Noting that $v=x+\frac{1}{x}$, $x^{\overline{\gamma}}=\frac{1}{x}$ and $l^{\overline{\gamma}}=l+\frac{1}{2}$, we finally test the analytic continuation \begin{equation}
    \bP_a=\mu^R_a{}^{\dot b}(\bP_{\dot{b}})^{\overline{\gamma}}.
\end{equation}
Following this algorithm to an appropriate order in $h$, the dimension of the state in consideration is fixed to the desired order.

\section{Comparison with the AdS$_3$ Virasoro-Shapiro Amplitude}\label{sec:comp-vs}

The states we study differ from those appearing in the Virasoro-Shapiro
amplitude approach~\cite{Chester:2024wnb,Jiang:2025oar,Jiang:2026xnh}. Following the AdS$_3$
integrability convention~\cite{Borsato:2013qpa}, we label multiplets by
the highest weight state obtained from the superconformal primary by
acting with two right-moving supercharges. Our states have
even spin $S\in2\mathbb{Z}_{>0}$ and equal left and right R-charges
$\frac{J}{2}$, so the corresponding superconformal primaries have
odd spin $S-1$ and R-charges $[\frac{J}{2},\frac{J-2}{2}]$. The
Virasoro-Shapiro amplitude instead exchanges multiplets whose
superconformal primaries have spin $l\in2\mathbb{Z}_{>0}$ and equal
left and right R-charges $\frac{p-1}{2}$ ($p\in\mathbb{Z}_{>0}$).



We note in particular that for $J=2$, the $(4,4)$ superconformal
primary is \textit{not} part of the lowest massive string multiplet. In contrast, in
$\mathcal{N}=4$ SYM theory and its string dual, the analogous state (meaning the $\mathcal{N}=4$
state with all Cartan charges outside $\mathfrak{psu}(1,1|2)^{\oplus 2}$
set to zero) would be a descendant of the Konishi multiplet, obtained
by the action of two supercharges. These supercharges, however, are not part the small $(4,4)$ superconformal algebra and are broken in the AdS$_3$ setting. As a result, the $J=2$
$(4,4)$ primary we study here sits in a distinct long multiplet
within the lowest massive string level, unrelated to the Konishi.

\section{Data}\label{SM:data}

In this appendix, we present our complete numerical results for the anomalous dimensions of operators with charges $S$ and $J$ for a range of couplings $h$. The numerical uncertainty below in $\Delta$ is of order $10^{-10}$.
\begin{table}[h]
\centering
\begin{tabular}{|cc|cc|cc|cc|cc|cc|cc|}
\hline
$h$ & $\Delta$ & $h$ & $\Delta$ & $h$ & $\Delta$ & $h$ & $\Delta$ & $h$ & $\Delta$ & $h$ & $\Delta$ & $h$ & $\Delta$ \\
\hline
\multicolumn{14}{|c|}{\textbf{$S = 2$, $J = 2$}} \\
\hline
0.01 & 4.001207282 & 0.08 & 4.078225448 & 0.6 & 6.135613947 & 1.3 & 8.501275846 & 2 & 10.35712566 & 2.7 & 11.93117562 & 3.4 & 13.32182153 \\
0.02 & 4.004853461 & 0.09 & 4.098743466 & 0.7 & 6.521109233 & 1.4 & 8.789868414 & 2.1 & 10.59615815 & 2.8 & 12.13953087 & 3.5 & 13.50886858 \\
0.03 & 4.010963707 & 0.1 & 4.121454961 & 0.8 & 6.887778530 & 1.5 & 9.069517300 & 2.2 & 10.82997795 & 2.9 & 12.34439471 & 3.6 & 13.69337328 \\
0.04 & 4.019547323 & 0.2 & 4.445163873 & 0.9 & 7.237562305 & 1.6 & 9.340984008 & 2.3 & 11.05890840 & 3 & 12.54593600 & 3.7 & 13.87543619 \\
0.05 & 4.030597147 & 0.3 & 4.865328931 & 1 & 7.572295971 & 1.7 & 9.604931396 & 2.4 & 11.28324113 & 3.1 & 12.74431048 & 3.8 & 14.05515142 \\
0.06 & 4.044089416 & 0.4 & 5.303718475 & 1.1 & 7.893597294 & 1.8 & 9.861939847 & 2.5 & 11.50324016 & 3.2 & 12.93966217 & 3.9 & 14.23260721 \\
0.07 & 4.059984063 & 0.5 & 5.729701596 & 1.2 & 8.202859617 & 1.9 & 10.11252044 & 2.6 & 11.71914541 & 3.3 & 13.13212453 & 4 & 14.40788643 \\
\hline
\hline
\multicolumn{14}{|c|}{\textbf{$S = 2$, $J = 3$}} \\
\hline
0.01 & 5.000802328 & 0.08 & 5.051403636 & 0.6 & 6.654157824 & 1.3 & 8.840353308 & 2 & 10.62588434 & 2.7 & 12.16018454 & 3.4 & 13.52451905 \\
0.02 & 5.003216499 & 0.09 & 5.064886187 & 0.7 & 6.997386434 & 1.4 & 9.115576817 & 2.1 & 10.85798831 & 2.8 & 12.36416158 & 3.5 & 13.70848916 \\
0.03 & 5.007248380 & 0.1 & 5.079840990 & 0.8 & 7.330203728 & 1.5 & 9.383283024 & 2.2 & 11.08537760 & 2.9 & 12.56488416 & 3.6 & 13.89005041 \\
0.04 & 5.012897096 & 0.2 & 5.298979111 & 0.9 & 7.652184950 & 1.6 & 9.643999153 & 2.3 & 11.30831910 & 3 & 12.76250067 & 3.7 & 14.06929402 \\
0.05 & 5.020154907 & 0.3 & 5.604364921 & 1 & 7.963600453 & 1.7 & 9.898203972 & 2.4 & 11.52705650 & 3.1 & 12.95714877 & 3.8 & 14.24630568 \\
0.06 & 5.029007215 & 0.4 & 5.948183326 & 1.1 & 8.265002365 & 1.8 & 10.14633103 & 2.5 & 11.74181285 & 3.2 & 13.14895642 & 3.9 & 14.42116599 \\
0.07 & 5.039432710 & 0.5 & 6.302548717 & 1.2 & 8.557035834 & 1.9 & 10.38877269 & 2.6 & 11.95279280 & 3.3 & 13.33804278 & 4 & 14.59395083 \\
\hline
\hline
\multicolumn{14}{|c|}{\textbf{$S = 2$, $J = 4$}} \\
\hline
0.01 & 6.000553648 & 0.08 & 6.035369982 & 0.6 & 7.296462457 & 1.3 & 9.284668920 & 2 & 10.98447065 & 2.7 & 12.46861871 & 3.4 & 13.79912532 \\
0.02 & 6.002217055 & 0.09 & 6.044672167 & 0.7 & 7.596197143 & 1.4 & 9.543901936 & 2.1 & 11.20791770 & 2.8 & 12.66700092 & 3.5 & 13.97910773 \\
0.03 & 6.004991668 & 0.1 & 6.055011825 & 0.8 & 7.892673594 & 1.5 & 9.797198337 & 2.2 & 11.42723992 & 2.9 & 12.86241884 & 3.6 & 14.15684814 \\
0.04 & 6.008875888 & 0.2 & 6.209863436 & 0.9 & 8.183883668 & 1.6 & 10.04484383 & 2.3 & 11.64264046 & 3 & 13.05499670 & 3.7 & 14.33242651 \\
0.05 & 6.013865044 & 0.3 & 6.436819010 & 1 & 8.468891898 & 1.7 & 10.28712374 & 2.4 & 11.85430849 & 3.1 & 13.24485074 & 3.8 & 14.50591827 \\
0.06 & 6.019951411 & 0.4 & 6.706488396 & 1.1 & 8.747349799 & 1.8 & 10.52431479 & 2.5 & 12.06242010 & 3.2 & 13.43208983 & 3.9 & 14.67739471 \\
0.07 & 6.027124265 & 0.5 & 6.997424134 & 1.2 & 9.019228184 & 1.9 & 10.75668080 & 2.6 & 12.26713925 & 3.3 & 13.61681606 & 4 & 14.84692319 \\
\hline
\hline
\multicolumn{14}{|c|}{\textbf{$S = 2$, $J = 5$}} \\
\hline
0.01 & 7.000400366 & 0.08 & 7.025562770 & 0.6 & 8.030782349 & 1.3 & 9.820243187 & 2 & 11.42470860 & 2.7 & 12.85095733 & 3.4 & 14.14159254 \\
0.02 & 7.001602433 & 0.09 & 7.032301387 & 0.7 & 8.289844066 & 1.4 & 10.06206318 & 2.1 & 11.63826858 & 2.8 & 13.04278945 & 3.5 & 14.31683326 \\
0.03 & 7.003606526 & 0.1 & 7.039802553 & 0.8 & 8.550953222 & 1.5 & 10.29951841 & 2.2 & 11.84833655 & 2.9 & 13.23198059 & 3.6 & 14.49002235 \\
0.04 & 7.006411457 & 0.2 & 7.153783104 & 0.9 & 8.811291648 & 1.6 & 10.53267532 & 2.3 & 12.05505113 & 3 & 13.41862927 & 3.7 & 14.66122738 \\
0.05 & 7.010014511 & 0.3 & 7.326580176 & 1 & 9.069171268 & 1.7 & 10.76163949 & 2.4 & 12.25854579 & 3.1 & 13.60282879 & 3.8 & 14.83051250 \\
0.06 & 7.014411466 & 0.4 & 7.539934509 & 1.1 & 9.323608373 & 1.8 & 10.98653903 & 2.5 & 12.45894823 & 3.2 & 13.78466748 & 3.9 & 14.99793867 \\
0.07 & 7.019596607 & 0.5 & 7.778372386 & 1.2 & 9.574056916 & 1.9 & 11.20751381 & 2.6 & 12.65638018 & 3.3 & 13.96422898 & 4 & 15.16356384 \\
\hline
\hline
\multicolumn{14}{|c|}{\textbf{$S = 2$, $J = 6$}} \\
\hline
0.01 & 8.000301382 & 0.08 & 8.019243955 & 0.6 & 8.831959867 & 1.3 & 10.43307282 & 2 & 11.93775134 & 2.7 & 13.30098096 & 3.4 & 14.54724543 \\
0.02 & 8.001205951 & 0.09 & 8.024325568 & 0.7 & 9.055016989 & 1.4 & 10.65706217 & 2.1 & 12.14067306 & 2.8 & 13.48557685 & 3.5 & 14.71715886 \\
0.03 & 8.002713729 & 0.1 & 8.029987823 & 0.8 & 9.283677887 & 1.5 & 10.87813753 & 2.2 & 12.34073697 & 2.9 & 13.66786938 & 3.6 & 14.88522468 \\
0.04 & 8.004823913 & 0.2 & 8.116857576 & 0.9 & 9.514874201 & 1.6 & 11.09618439 & 2.3 & 12.53802126 & 3 & 13.84793159 & 3.7 & 15.05149775 \\
0.05 & 8.007534877 & 0.3 & 8.251609095 & 1 & 9.746552049 & 1.7 & 11.31115569 & 2.4 & 12.73260665 & 3.1 & 14.02583390 & 3.8 & 15.21603059 \\
0.06 & 8.010844175 & 0.4 & 8.422640984 & 1.1 & 9.977351693 & 1.8 & 11.52305042 & 2.5 & 12.92457460 & 3.2 & 14.20164405 & 3.9 & 15.37887349 \\
0.07 & 8.014748550 & 0.5 & 8.619017595 & 1.2 & 10.20638302 & 1.9 & 11.73189866 & 2.6 & 13.11400614 & 3.3 & 14.37542709 & 4 & 15.54007461 \\
\hline
\hline
\multicolumn{14}{|c|}{\textbf{$S = 2$, $J = 7$}} \\
\hline
0.02 & 9.000937819 & 0.04 & 9.003750964 & 0.06 & 9.008433210 & 0.08 & 9.014970705 & 0.1 & 9.023342168 &  &  &  &  \\
0.03 & 9.002110185 & 0.05 & 9.005859138 & 0.07 & 9.011471210 & 0.09 & 9.018928802 &  &  &  &  &  &  \\
\hline
\hline
\multicolumn{14}{|c|}{\textbf{$S = 2$, $J = 8$}} \\
\hline
0.05 & 10.00467917 & 0.06 & 10.00673536 & 0.07 & 10.00916275 & 0.08 & 10.01195982 & 0.09 & 10.01512472 & 0.1 & 10.01865537 &  &  \\
\hline\end{tabular}
\caption{Dimension of operators for various states}
\label{tab:allstates}
\end{table} \begin{table}[h]
\centering
\begin{tabular}{|cc|cc|cc|cc|cc|cc|cc|}
\hline
$h$ & $\Delta$ & $h$ & $\Delta$ & $h$ & $\Delta$ & $h$ & $\Delta$ & $h$ & $\Delta$ & $h$ & $\Delta$ & $h$ & $\Delta$ \\
\hline
\multicolumn{14}{|c|}{\textbf{$S = 4$, $J = 2$}} \\
\hline
0.1 & 6.171427121 & 0.7 & 9.535460935 & 1.3 & 12.29122301 & 1.9 & 14.53884762 & 2.5 & 16.48317513 & 3.1 & 18.22119771 & 3.7 & 19.80726508 \\
0.2 & 6.630181806 & 0.8 & 10.04552155 & 1.4 & 12.69338457 & 2 & 14.88054761 & 2.6 & 16.78534988 & 3.2 & 18.49499258 & 3.8 & 20.05941816 \\
0.3 & 7.223132199 & 0.9 & 10.53207751 & 1.5 & 13.08325462 & 2.1 & 15.21458042 & 2.7 & 17.08217943 & 3.3 & 18.76479199 & 3.9 & 20.30844005 \\
0.4 & 7.838282371 & 1 & 10.99779223 & 1.6 & 13.46188103 & 2.2 & 15.54143851 & 2.8 & 17.37393756 & 3.4 & 19.03076589 & 4 & 20.55444461 \\
0.5 & 8.433369477 & 1.1 & 11.44496598 & 1.7 & 13.83017406 & 2.3 & 15.86156407 & 2.9 & 17.66087551 & 3.5 & 19.29307255 &  &  \\
0.6 & 8.998967873 & 1.2 & 11.87555563 & 1.8 & 14.18892956 & 2.4 & 16.17535591 & 3 & 17.94322445 & 3.6 & 19.55185961 &  &  \\
\hline
\hline
\multicolumn{14}{|c|}{\textbf{$S = 4$, $J = 3$}} \\
\hline
0.1 & 7.120966473 & 0.7 & 9.933697089 & 1.3 & 12.55892384 & 1.9 & 14.75142637 & 2.5 & 16.66404565 & 3.1 & 18.38097531 & 3.7 & 19.95176975 \\
0.2 & 7.452503142 & 0.8 & 10.41023653 & 1.4 & 12.94911533 & 2 & 15.08674094 & 2.6 & 16.96214856 & 3.2 & 18.65190032 & 3.8 & 20.20176659 \\
0.3 & 7.908837693 & 0.9 & 10.86990720 & 1.5 & 13.32840298 & 2.1 & 15.41490513 & 2.7 & 17.25515859 & 3.3 & 18.91897371 & 3.9 & 20.44872289 \\
0.4 & 8.415492935 & 1 & 11.31352115 & 1.6 & 13.69759199 & 2.2 & 15.73634659 & 2.8 & 17.54332493 & 3.4 & 19.18235401 & 4 & 20.69274641 \\
0.5 & 8.931925714 & 1.1 & 11.74215628 & 1.7 & 14.05740467 & 2.3 & 16.05145346 & 2.9 & 17.82687745 & 3.5 & 19.44218924 &  &  \\
0.6 & 9.440265336 & 1.2 & 12.15693336 & 1.8 & 14.40848884 & 2.4 & 16.36057900 & 3 & 18.10602861 & 3.6 & 19.69861790 &  &  \\
\hline
\hline
\multicolumn{14}{|c|}{\textbf{$S = 4$, $J = 4$}} \\
\hline
0.01 & 8.000878797 & 0.08 & 8.056391594 & 0.6 & 9.996709987 & 1.3 & 12.91338223 & 2 & 15.36351991 & 2.7 & 17.48899926 & 3.4 & 19.38819827 \\
0.02 & 8.003522309 & 0.09 & 8.071240479 & 0.7 & 10.44298068 & 1.4 & 13.28865635 & 2.1 & 15.68414547 & 2.8 & 17.77248319 & 3.5 & 19.64478197 \\
0.03 & 8.007936951 & 0.1 & 8.087743676 & 0.8 & 10.88132504 & 1.5 & 13.65467247 & 2.2 & 15.99861037 & 2.9 & 18.05161545 & 3.6 & 19.89810453 \\
0.04 & 8.014123318 & 0.2 & 8.333809063 & 0.9 & 11.30951572 & 1.6 & 14.01196194 & 2.3 & 16.30723744 & 3 & 18.32658527 & 3.7 & 20.14828534 \\
0.05 & 8.022076090 & 0.3 & 8.689514097 & 1 & 11.72674009 & 1.7 & 14.36103169 & 2.4 & 16.61032468 & 3.1 & 18.59756906 & 3.8 & 20.39543679 \\
0.06 & 8.031784032 & 0.4 & 9.105309717 & 1.1 & 12.13291167 & 1.8 & 14.70235789 & 2.5 & 16.90814740 & 3.2 & 18.86473149 & 3.9 & 20.63966483 \\
0.07 & 8.043230095 & 0.5 & 9.547461271 & 1.2 & 12.52830852 & 1.9 & 15.03638379 & 2.6 & 17.20096028 & 3.3 & 19.12822650 & 4 & 20.88106951 \\
\hline
\hline
\multicolumn{14}{|c|}{\textbf{$S = 6$, $J = 2$}} \\
\hline
0.1 & 8.203569714 & 0.7 & 12.20603873 & 1.3 & 15.49897496 & 1.9 & 18.19885941 & 2.5 & 20.54264637 & 3.1 & 22.64287866 & 3.7 & 24.56293047 \\
0.2 & 8.749805252 & 0.8 & 12.81398438 & 1.4 & 15.98120006 & 2 & 18.61026177 & 2.6 & 20.90747748 & 3.2 & 22.97411099 & 3.8 & 24.86844381 \\
0.3 & 9.455349474 & 0.9 & 13.39455794 & 1.5 & 16.44906143 & 2.1 & 19.01264899 & 2.7 & 21.26599112 & 3.3 & 23.30060162 & 3.9 & 25.17022884 \\
0.4 & 10.18687108 & 1 & 13.95086879 & 1.6 & 16.90376973 & 2.2 & 19.40659215 & 2.8 & 21.61850638 & 3.4 & 23.62254953 & 4 & 25.46841936 \\
0.5 & 10.89444030 & 1.1 & 14.48558849 & 1.7 & 17.34637701 & 2.3 & 19.79260451 & 2.9 & 21.96531626 & 3.5 & 23.94014012 &  &  \\
0.6 & 11.56726809 & 1.2 & 15.00098332 & 1.8 & 17.77780355 & 2.4 & 20.17114932 & 3 & 22.30669063 & 3.6 & 24.25354647 &  &  \\
\hline
\hline
\multicolumn{14}{|c|}{\textbf{$S = 8$, $J = 2$}} \\
\hline
0.01 & 10.00219865 & 0.05 & 10.05662485 & 0.5 & 13.24020821 & 0.9 & 16.04886720 & 1.3 & 18.42319871 & 1.7 & 20.51454711 & 2.1 & 22.40555577 \\
0.02 & 10.00888017 & 0.2 & 10.83871352 & 0.6 & 13.99481542 & 1 & 16.67564783 & 1.4 & 18.96857094 & 1.8 & 21.00376124 & 2.2 & 22.85321150 \\
0.03 & 10.02014469 & 0.3 & 11.62834664 & 0.7 & 14.71204219 & 1.1 & 17.27871723 & 1.5 & 19.49801146 & 1.9 & 21.48149240 &  &  \\
0.04 & 10.03605291 & 0.4 & 12.44740724 & 0.8 & 15.39546403 & 1.2 & 17.86053854 & 1.6 & 20.01296530 & 2 & 21.94852448 &  &  \\
\hline\end{tabular}
\caption{Dimension of operators for various states}
\label{tab:allstates}
\end{table}
\end{document}